\documentclass[12pt]{article}

\begin{document}

\title{\bf \Large Energy and Momentum in Spacetime Homogeneous
G$\ddot{o}$del-type Metrics}

\author{M. Sharif \thanks{Present Address: Department of Mathematical Sciences,
University of Aberdeen, Kings College, Aberdeen AB24 3UE Scotland,
UK. $<$msharif@maths.abdn.ac.uk$>$}
\\ Department of Mathematics, University of the Punjab,\\
Quaid-e-Azam Campus Lahore-54590, PAKISTAN,\\
$<$hasharif@yahoo.com$>$}

\date{}

\maketitle

\begin{abstract}
Using Einstein and Papapetrou energy-momentum complexes, we
explicitly calculate the energy and momentum distribution
associated with spacetime homogeneous G$\ddot{o}$del-type metrics.
We obtain that the two definitions of energy-momentum complexes do
not provide the same result for these type of metrics. However, it
is shown that the results obtained are reduced to the
energy-momentum densities of G$\ddot{o}$del metric already
available in the literature.
\end{abstract}
{\bf Key Words: Energy and Momentum, G$\ddot{o}$del-type Metrics}\\
{\bf PACS: 04.20.Cv}

\newpage

\section{Introduction}

The problem of energy and momentum has been one of the oldest but
most interesting problems in Einstein's theory of General
Relativity (GR). Due to its peculiar nature and diverse points of
view, it has been the most controversial problem. In a curved
spacetime the energy-momentum tensor of matter plus all
non-gravitational fields no longer satisfies $T^b_{a;b}=0$. The
contribution from the gravitational field is now required to
construct an energy-momentum expression which satisfies a
divergence relation. Einstein himself started work to solve this
problem and suggested an expression for energy-momentum
distribution [1]. He justified that his energy-momentum complex
provides convincing results for the total energy and momentum of
isolated systems. After this, many physicists including
Landau-Lifshitz [2], Tolman [3], Papapetrou [4], Bergmann [5],
Weinberg [6] had suggested different expressions for the
energy-momentum distribution. The main problem with these
definitions is that they are coordinate dependent. One can have
meaningful results only when calculations are performed in
Cartesian coordinates. This restriction of coordinate dependent
motivated some other physicists like M$\ddot{o}$ller [7]-[8],
Komar [9] and Penrose [10] who constructed coordinate independent
definitions of energy-momentum complex.

M$\ddot{o}$ller claimed that his expression gives the same values
for the total energy and momentum as the Einstein's
energy-momentum complex for a closed system. However,
M$\ddot{o}$ller's energy-momentum complex was subjected to some
criticism [8]-[11]. Komar's prescription, though not restricted to
the use of Cartesian coordinates, is not applicable to non-static
space-times. Penrose [10] pointed out that quasi-local masses are
conceptually very important. The inadequacies of these quasi-local
masses (these different definitions do not give agreed results for
the Reissner- Nordstrom and Kerr metrics and that the Penrose
definition could not succeed to deal with the Kerr metric) have
been discussed in [12]-[14]. Thus each of these energy-momentum
complex has its own drawback. As a result these ideas of the
energy-momentum complex were severally criticized.

Virbhadra [14]-[15] was the first who revived the interest in this
approach. Since then lot of work on evaluating the energy-momentum
distributions of different spacetimes have been carried out by
different authors [16]-[19]. In a recent paper, Virbhadhra [14]
investigated whether or not the energy-momentum complexes of
Einstein, Landau and Lifshitz, Papapetrou and Weinberg give the
same energy distribution for the most general non-static
spherically symmetric metric. It was a great surprise that
contrary to previous results of many asymptotically flat
spacetimes and asymptotically non-flat spacetimes, he found that
these definitions disagree. He observed that Einstein's
energy-momentum complex provides a consistent result for the
Schwarzschild metric whether one calculates in Kerr-Schild
Cartesian coordinates or Schwarzschild Cartesian coordinates. The
prescriptions of Landau-Lifshitz, Papapetrou and Weinberg furnish
the same result as in the Einstein prescription if the
calculations are carried out in Schwarzschild Cartesian
coordinates. Thus the prescriptions of Landau-Lifshitz, Papapetrou
and Weinberg do not give a consistent result. On the basis of
these and some other facts [12,13], Virbhadra concluded that the
Einstein method seems to be the best among all known (including
quasi-local mass definitions) for energy distribution in a
spacetime. Recently, Lessner [20] pointed out that the
M$\ddot{o}$ller's energy-momentum prescription is a powerful
concept of energy and momentum in GR.

In a series of papers [21] Cooperstock has propounded a hypothesis
according to which, in a curved spacetime, energy and momentum are
confined to the regions of non-vanishing energy-momentum tensor
$T_a^b$ of the matter and all non-gravitational fields. The
results of Xulu [22] and the recent results of Bringley [23]
support this hypothesis. It would be interesting to investigate
further whether or not the Cooperstock's hypothesis stands true.
In a recent paper [19], we have applied Einstein and Papapetrou's
prescriptions to calculate energy-momentum densities of
G$\ddot{o}$del spacetime. The results obtained for the
G$\ddot{o}$del metric are much simple in both the prescriptions.
We obtain that the energy density is exactly the same in both the
prescriptions with the exception of different signs in the first
term while the momentum density components exactly coincide up to
the first term only.

In this paper we extend the procedure and use Einstein and
Papapetrou's prescriptions to evaluate energy and momentum
densities in G$\ddot{o}$del-type metrics. As we shall see from the
analysis given in the paper, when procedure is extended to such
metrics, the problem becomes considerably complicated. We find
that the results obtained by these two prescriptions are not the
same. However, it is shown that they both reduce to the known
results for particular values of $H$ and $D$, i.e., G$\ddot{o}$del
spacetime. In the next section, we shall describe the spacetime
homogeneous G$\ddot{o}$del-type metrics. In sections three and
four, we evaluate energy and momentum using Einstein and
Papapetrou's prescriptions respectively. Finally, the results
obtained will be concluded.

\section{Spacetime Homogeneous G$\ddot{o}$del-type Metrics}

A solution of Einstein's field equations with cosmological
constant for incoherent matter with rotation was found by
G$\ddot{o}$del. This is the best known example of a cosmological
model which makes it apparent that GR does not exclude the
existence of a closed timelike world-lines, despite its Lorentzian
character which leads to the local validity of the causality
principle. G$\ddot{o}$del-type metrics are given by the line
element of the form [24]
\begin{equation}
ds^2=[dt+H(r)d\theta]^2-dr^2-D^2(r)d\theta^2-dz^2,
\end{equation}
in the cylindrical coordinates $(r,\theta,z)$. Here the metric
functions $H$ and $D$ depend on the coordinate $r$ only. It admits
a five-parameter group of isometries ($G_5$) having an isotropy
subgroup of dimension one ($H_1$).

Raychaudhuri and Thakurta [25] are the first who have determined
the necessary conditions for a G$\ddot{o}$del-type metric to be a
spacetime homogeneous (hereafter called ST homogeneous). Later,
Reboucas and Tiomno [26] proved that these conditions are also
sufficient for ST homogeneity of G$\ddot{o}$del-type Riemannian
spacetime manifolds. These necessary and sufficient conditions are
given by
\begin{equation}
\frac{D''}{D}=constant\equiv m^2,
\end{equation}
\begin{equation}
\frac{H'}{D}=constant\equiv -2\omega.
\end{equation}
The necessary and sufficient conditions were finally re-derived
for a G$\ddot{o}$del-type manifold to be ST homogeneous without
assuming any such simplifying hypothesis in [27].

We can distinguish the ST metrics in the following four classes as
given in [28-29] according to\\
\par \noindent
\par \noindent
{\bf Class I}: $m^2>0,~\omega\neq 0$. In this case, the general
solution of Eqs.(2) and (3) is given by
\begin{equation}
H(r)=\frac{2\omega}{m^2}[1-\cosh(mr)],\quad \quad
D(r)=\frac{1}{m}\sinh(mr).
\end{equation}
{\bf Class II}: $m^2=0,~\omega\neq 0$. For this class, the general
solution of Eqs.(2) and (3) can be written as
\begin{equation}
H(r)=-\omega r^2,\quad \quad D(r)=r.
\end{equation}
{\bf Class III}: $m^2\equiv -\mu^2,~\omega\neq 0$. If we integrate
Eqs.(2) and (3) for this case, we have the following solution
\begin{equation}
H(r)=\frac{2\omega}{\mu^2}[\cos(\mu r)-1],\quad \quad
D(r)=\frac{1}{\mu}\sin(\mu r).
\end{equation}
{\bf Class IV}: $m^2\neq 0,~\omega=0$. In this case, the cross
term related to the rotation $\omega$ in the G$\ddot{o}$del model
vanishes. Consequently, one can make $H=0$ by a trivial coordinate
transformation.

If $m^2=0=\omega$, the line element (1) becomes Minkowskian. Also,
it is mentioned that the case $m^2=2\omega^2$ defines the original
G$\ddot{o}$del metric.

In order to have meaningful results in the prescriptions of
Einstein and Papapetrou, it is necessary to transform the metric
in Cartesian coordinates. We transform the metric in Cartesian
coordinates by using
\begin{equation}
x=r\cos\theta,\quad y=r\sin\theta.
\end{equation}
The corresponding metric in these coordinates will become
\begin{equation}
ds^2=dt^2-\frac{1}{r^2}(xdx+ydy)^2+\frac{1}{r^4}(H^2-D^2)(xdy-ydx)^2
-dz^2+\frac{2}{r^2}Hdt(xdy-ydx).
\end{equation}

\section{Energy and Momentum in Einstein's Prescription}

The energy-momentum complex of Einstein [1] is given by
\begin{equation}
\Theta_a^b=\frac{1}{16\pi}M^{bc}_{a\;,c},
\end{equation}
where
\begin{equation}
M_a^{bc}=\frac{g_{ad}}{\sqrt{-g}}[-g(g^{bd}g^{ce}-g^{cd}g^{be})]_{,e},\quad
a,b,c,d,e=0,1,2,3.
\end{equation}
$\Theta_0^0$ is the energy density, $\Theta_0^a$ are the momentum
density components, and $\Theta_a^0$ are the components of energy
current density. The Einstein energy-momentum satisfies the local
conservation laws
\begin{equation}
\frac{\partial\Theta_a^b}{\partial x^b}=0.
\end{equation}
In order to evaluate the energy and momentum densities in
Einstein's prescription associated with G$\ddot{o}$del-type
metrics, we need to calculate the non-vanishing components of
$M_a^{bc}$
\begin{equation}
M_0^{01}=\frac{1}{Dr^3}(D^2x+HH_1x^2+HH_2xy-2DD_1x^2-2DD_2xy+r^2x),
\end{equation}
\begin{eqnarray}
M_1^{01}=-\frac{1}{Dr^5}(H^2H_1x^2y+H^2H_2xy^2-2HDD_1x^2y
-2HDD_2xy^2\nonumber\\
-H_1r^2x^2y+H_2r^2x^3+D^2H_1x^2y+D^2H_2xy^2),
\end{eqnarray}
\begin{eqnarray}
M_1^{02}=-\frac{1}{Dr^5}(Hr^4+H^2H_2y^3+H^2H_1xy^2-2HDD_2y^3
-2HDD_1xy^2\nonumber\\
+H_2r^2x^2y-H_1r^2xy^2+D^2H_2y^3+D^2H_1xy^2),
\end{eqnarray}
\begin{eqnarray}
M_2^{01}=\frac{1}{Dr^5}(Hr^4+H^2H_1x^3+H^2H_2x^2y-2HDD_1x^3
-2HDD_2x^2y\nonumber\\
+H_1r^2xy^2-H_2r^2x^2y+D^2H_1x^3+D^2H_2x^2y),
\end{eqnarray}
\begin{eqnarray}
M_2^{02}=\frac{1}{Dr^5}(H^2H_1x^2y+H^2H_2xy^2-2HDD_1x^2y
-2HDD_2xy^2\nonumber\\
+H_1r^2y^3-H_2r^2xy^2+D^2H_1x^2y+D^2H_2xy^2),
\end{eqnarray}
\begin{equation}
M_3^{03}=\frac{1}{Dr}(H_1y-H_2x),
\end{equation}
\begin{equation}
M_0^{12}=\frac{1}{Dr}(H_1x+H_2y).
\end{equation}
It is to be noted that $H_1$ and $H_2$ denote differentiation of
$H$ with respect to the coordinates $x$ and $y$ respectively.
Using Eqs.(12)-(18) in Eq.(9), we obtain the energy and momentum
densities in Einstein's prescription
\begin{eqnarray}
\Theta_0^0=\frac{1}{16\pi
D^2r^3}(-D^3+D^2D_1x+D^2D_2y-2D^2D_{11}x^2-2D^2D_{22}y^2\nonumber\\
-4D^2D_{12}xy+Dr^2-D_1r^2x-D_2r^2y+HH_{11}Dx^2+HH_{22}Dy^2\nonumber\\
+2HH_{12}Dxy+2H_1H_2Dxy+H_1^2Dx^2+H_2^2Dy^2\nonumber\\
-HH_1D_1x^2-HH_2D_1xy-HH_1D_2xy-HH_2D_2y^2),
\end{eqnarray}
\begin{eqnarray}
\Theta_1^0=\frac{1}{16\pi
D^2r^5}(H^2H_1Dxy+H^2H_2Dy^2-H^2H_{11}Dx^2y-H^2H_{22}Dy^3\nonumber\\
-2H^2H_{12}Dxy^2+H^2H_1D_1x^2y+H^2H_1D_2xy^2+H^2H_2D_1xy^2\nonumber\\
+H^2H_2D_2y^3-2HH_1^2Dx^2y-2HH_2^2Dy^3-4HH_1H_2Dxy^2\nonumber\\
-2HD^2D_1xy-2HD^2D_2y^2+2HD^2D_{11}x^2y+2HD^2D_{22}y^3\nonumber\\
+4HD^2D_{12}xy^2+D^3H_1xy+D^3H_2y^2-D^3H_{11}x^2y-D^3H_{22}y^3\nonumber\\
-2D^3H_{12}xy^2+D^2D_1H_1x^2y+D^2D_1H_2xy^2+D^2D_2H_1xy^2\nonumber\\
+D^2D_2H_2y^3+HDr^2y+HD_2r^4+DH_1r^2xy-DH_2r^2x^2\nonumber\\
-DH_2r^4+DH_{11}r^2x^2y-DH_{22}r^2x^2y+DH_{12}r^2xy^2\nonumber\\
-DH_{12}r^2x^3-H_1D_1r^2x^2y-H_1D_2r^2xy^2+H_2D_1r^2x^3+H_2D_2r^2x^2y),
\end{eqnarray}
\begin{eqnarray}
\Theta_2^0=-\frac{1}{16\pi
D^2r^5}(H^2H_2Dxy+H^2H_1Dx^2-H^2H_{22}Dxy^2-H^2H_{11}Dx^3\nonumber\\
-2H^2H_{12}Dx^2y+H^2H_2D_2xy^2+H^2H_2D_1x^2y+H^2H_1D_2x^2y\nonumber\\
+H^2H_1D_1x^3-2HH_2^2Dxy^2-2HH_1^2Dx^3-4HH_1H_2Dx^2y\nonumber\\
-2HD^2D_2xy-2HD^2D_1x^2+2HD^2D_{22}xy^2+2HD^2D_{11}x^3\nonumber\\
+4HD^2D_{12}x^2y+D^3H_2xy+D^3H_1x^2-D^3H_{22}xy^2-D^3H_{11}x^3\nonumber\\
-2D^3H_{12}x^2y+D^2D_2H_2xy^2+D^2D_2H_1x^2y+D^2D_1H_2x^2y\nonumber\\
+D^2D_1H_1x^3+HDr^2x+HD_1r^4+DH_2r^2xy-DH_1r^2y^2\nonumber\\
-DH_1r^4+DH_{22}r^2xy^2-DH_{11}r^2xy^2+DH_{12}r^2x^2y\nonumber\\
-DH_{12}r^2y^3-H_2D_2r^2xy^2-H_2D_1r^2x^2y+H_1D_2r^2y^3+H_1D_1r^2xy^2),
\end{eqnarray}
\begin{eqnarray}
\Theta_0^1=\frac{1}{16\pi
D^2r^3}(DH_2r^2-DH_1xy-DH_2y^2+H_{12}r^2x+DH_{22}r^2y\nonumber\\
-D_2H_1r^2x-D_2H_2r^2y),
\end{eqnarray}
\begin{equation}
\Theta_0^2=-\Theta_0^1,
\end{equation}
\begin{equation}
\Theta_0^3=\Theta_3^0=0.
\end{equation}
Now for $H=e^{ar}$ and $D=e^{ar}/\sqrt{2}$, Eqs.(19)-(24) become
\begin{equation}
\Theta_0^0=\frac{(1-ar)}{16\sqrt{2}\pi r^3}[-e^{ar}+2r^2e^{-ar}],
\end{equation}
\begin{equation}
\Theta_1^0=\frac{y}{16\sqrt{2}\pi r^4}[2r+a(1-2ar)e^{2ar}],
\end{equation}
\begin{equation}
\Theta_2^0=-\frac{x}{16\sqrt{2}\pi r^4}[2r+a(1-2ar)e^{2ar}],
\end{equation}
\begin{equation}
\Theta_0^1=\Theta_0^2=\Theta_0^3=\Theta_3^0=0.
\end{equation}
These are the energy and momentum densities of G$\ddot{o}$del
spacetime given by Sharif [19].

\section{Energy and Momentum in Papapetrou's Prescription}

The symmetric energy-momentum complex of Papapetrou [4] is given
by
\begin{equation}
\Omega^{ab}=\frac{1}{16\pi}N^{abcd}_{\quad\;\,,cd},
\end{equation}
where
\begin{equation}
N^{abcd}=\sqrt{-g}(g^{ab}\eta^{cd}-g^{ac}\eta^{bd}
+g^{cd}\eta^{ab}-g^{bd}\eta^{ac}),
\end{equation}
and $\eta^{ab}$ is the Minkowski spacetime. The energy-momentum
complex satisfies the local conservation laws
\begin{equation}
\frac{\partial\Omega^{ab}}{\partial x^b}=0.
\end{equation}
The locally conserved energy-momentum complex $\Omega^{ab}$
contains contributions from the matter, non-gravitational and
gravitational fields. $\Omega^{00}$ and $\Omega^{0a}$ are the
energy and momentum (energy current) density components. To find
the energy and momentum densities of the spacetime under
consideration, we require the following non-zero components of
$N^{abcd}$ given as
\begin{equation}
N^{0011}=\frac{1}{Dr^3}(H^2r^2-D^2r^2-D^2x^2-r^2y^2),
\end{equation}
\begin{equation}
N^{0012}=\frac{1}{Dr^3}xy(r^2-D^2)=N^{0021},
\end{equation}
\begin{equation}
N^{0022}=\frac{1}{Dr^3}(H^2r^2-D^2r^2-D^2y^2-r^2x^2),
\end{equation}
\begin{equation}
N^{0121}=-\frac{1}{Dr}Hx,
\end{equation}
\begin{equation}
N^{0122}=-\frac{1}{Dr}Hy,
\end{equation}
\begin{equation}
N^{0211}=-N^{0121},
\end{equation}
\begin{equation}
N^{0212}=-N^{0122},
\end{equation}
Substituting Eqs.(32)-(38) in Eq.(29), we obtain the following
energy and momentum density components in Papapetrou's
prescription
\begin{eqnarray}
\Omega^{00}=\frac{1}{16\pi D^3r^3}(H^2D^2+2H^2DD_1x
+2H^2DD_2y+2H^2D_1^2r^2\nonumber\\
+2H^2D_2^2r^2-H^2DD_{11}r^2-H^2DD_{22}r^2-4HH_1D^2x\nonumber\\
-4HH_2D^2y-4HH_1DD_1r^2-4HH_2DD_2r^2+2HH_{11}D^2r^2\nonumber\\
+2HH_{22}D^2r^2+2H_1^2D^2r^2+2H_2^2D^2r^2-D^4+2D^3D_1x\nonumber\\
+2D^3D_2y-D^3D_{11}r^2-D^3D_{22}r^2-D^3D_{11}x^2-D^3D_{22}y^2\nonumber\\
-2D^3D_{12}xy+D^2r^2-2DD_1r^2x-2DD_2r^2y+DD_{11}r^2y^2\nonumber\\
+DD_{22}r^2x^2-2DD_{12}r^2xy-2D_1^2r^2y^2-2D_2^2r^2x^2+4D_1D_2r^2xy),
\end{eqnarray}
\begin{eqnarray}
\Omega^{01}=\frac{1}{16\pi D^3r^3}(HD^2y-HDD_1xy
+2HDD_2x^2+HDD_2y^2\nonumber\\
+HDD_{12}r^2x+HDD_{22}r^2y-2HD_1D_2r^2x-2HD_2^2r^2y\nonumber\\
+D^2H_1xy-2D^2H_2x^2-D^2H_2y^2-D^2H_{12}r^2x\nonumber\\
-D^2H_{22}r^2y+DD_1H_2r^2x+DD_2H_1r^2x+2DD_2H_2r^2y,
\end{eqnarray}
\begin{eqnarray}
\Omega^{02}=-\frac{1}{16\pi D^3r^3}(HD^2x-HDD_2xy
+2HDD_1y^2+HDD_1x^2\nonumber\\
+HDD_{12}r^2y+HDD_{11}r^2x-2HD_1D_2r^2y-2HD_1^2r^2x\nonumber\\
+D^2H_2xy-2D^2H_1y^2-D^2H_1x^2-D^2H_{12}r^2y\nonumber\\
-D^2H_{11}r^2x+DD_2H_1r^2y+DD_1H_2r^2y+2DD_1H_1r^2x,
\end{eqnarray}
\begin{equation}
\Omega^{03}=\Omega^{30}=0.
\end{equation}
We see that for $H=e^{ar}$ and $D=e^{ar}/\sqrt{2}$, Eqs.(39)-(42)
yield
\begin{equation}
\Omega^{00}=\frac{(1-ar)}{16\sqrt{2}\pi r^3}(e^{ar}+2r^2e^{-ar}),
\end{equation}
\begin{equation}
\Omega^{01}=\frac{y}{8\sqrt{2}\pi r^3},
\end{equation}
\begin{equation}
\Omega^{02}=-\frac{x}{8\sqrt{2}\pi r^3},
\end{equation}
\begin{equation}
\Omega^{03}=\Omega^{30}=0.
\end{equation}

These turn out to be the energy and momentum density components
for G$\ddot{o}$del spacetime given by Sharif [19].

\section{Discussion}

It has been remained a controversial problem whether energy and
momentum are localizable or not. Misner et al [30] were the points
of view that energy can only be localized for spherical systems.
Cooperstock and Sarracino [31] argued that if energy can be
localized in spherical systems then it can be localized in any
spacetimes. Bondi [32] supported the point of view that a
non-localizable form of energy cannot be allowed in GR and hence
its localization can be found in principle. The energy-momentum
complexes are non-tensorial under general coordinate
transformations and hence are restricted to Cartesian coordinates
only. Virbhadra and others have shown [14]-[19] that these
energy-momentum complexes can provide meaningful results.

In this paper, we have evaluated the energy and momentum density
components for G$\ddot{o}$del-type metrics by using prescriptions
of Einstein and Papapetrou. It can be seen that the energy and
momentum densities turn out to be finite and well defined in both
the prescriptions. These provide general results in terms of $H$
and $D$ which can furnish interesting results for special values
of $H$ and $D$. It follows from Eqs.(19)-(24) and (39)-(42) that
the two results obtained by using the Einstein and Papapetrou
energy-momentum complex differ in general for G$\ddot{o}$del-type
metrics. This should be considered important why the two results
are different. It is to be noted from Eqs.(25)-(28) and (43)-(46)
that both the results reduce to the known energy and momentum
densities of a G$\ddot{o}$del metric as given in [19].

There are spacetimes [14], [19], [33] for which the two or more
energy momentum complexes do not give the same result. We have
exposed another model for which the two energy-momentum complexes
do not provide the consistent result. This is another example
which indicate that the idea of localization does not follow along
the lines of pseudo-tensorial construction but instead it follows
from the energy-momentum tensor itself.

\newpage

\begin{description}
\item  {\bf Acknowledgment}
\end{description}

The author would like to thank Higher Education Commission (HEC),
Pakistan for providing postdoctoral fellowship at University of
Aberdeen, UK. I am also grateful for the useful comments made by
the anonymous referee.

\vspace{2cm}

{\bf \large References}

\begin{description}

\item{[1]} Trautman, A.: {\it Gravitation: An Introduction to
Current Research} ed. Witten, L. (Wiley, New York, 1962)169.

\item{[2]} Landau, L.D. and Lifshitz, E.M.: {\it The Classical
Theory of Fields} (Addison-Wesley Press, Reading, MA, 1962)2nd ed.

\item{[3]} Tolman, R.C.: {\it Relativity, Thermodynamics and
Cosmology} (Oxford Univ. Press, 1934)227.

\item{[4]} Papapetrou, A.: Proc. R. Irish. Acad. {\bf
A52}(1948)11.

\item{[5]} Bergmann, P.G. and Thompson, R.: Phys. Rev. {\bf
89}(1953)400.

\item{[6]} Weinberg, S.: {\it Gravitation and Cosmology} (Wiley,
New York, 1972).

\item{[7]} M$\ddot{o}$ller, C.: Ann. Phys. (NY) {\bf 4}(1958)347.

\item{[8]} M$\ddot{o}$ller, C.: Ann. Phys. (NY) {\bf 12}(1961)118.

\item{[9]} Komar, A.: Phys. Rev. {\bf 113}(1959)934.

\item{[10]} Penrose, R.: Proc. Roy. Soc. London {\bf
A381}(1982)53.

\item{[11]} Kovacs, D.: Gen. Rel. and Gravit. {\bf 17}(1985)927;

Novotny, J.: Gen. Rel. and Gravit. {\bf 19}(1987)1043;

\item{[12]} Bergqvist, G.: Class. Quantum Gravit. {\bf
9}(1992)1753.

\item{[13]} Bernstein, D.H. and Tod, K.P.: Phys. Rev. {\bf
D49}(1994)2808.

\item{[14]} Virbhadra, K.S.: Phys. Rev. {\bf D60}(1999)104041.

\item{[15]} Virbhadra, K.S.: Phys. Rev. {\bf D41}(1990)1081; {\bf
D42}(1990)1066;\\
{\bf D42}(1990)2919; and references therein.

\item{[16]} Xulu, S.S.: Int. J. Mod. Phys. {\bf A15}(2000)2979;
Mod. Phys. Lett. {\bf A15}(2000)1511 and references therein.

\item{[17]} Yang, I.C. and Radinschi, I.: Mod. Phys. Lett. {\bf
A17}(2002)1159.

\item{[18]} Sharif, M.: Int. J. of Mod. Phys. {\bf A17}(2002)1175.

\item{[19]} Sharif, M.: Int. J. of Mod. Phys. {\bf A18}(2003);
Erratum {\bf A}(2003).

\item{[20]} Lessner, G.: Gen. Rel. Grav. {\bf 28}(1996)527.

\item{[21]} Cooperstock, F.I.: in {\it Topics on Quantum Gravity
and Beyond}, Essays in honour of Witten, L. on his retirement, ed.
Mansouri, F. and Scanio, J.J. (World Scientific, Singapore, 1993);
Mod. Phys. Lett. {\bf A14}(1999)1531; Annals of Phys. {\bf
282}(2000)115;\\
Cooperstock, F.I. and Tieu, S.: Found. Phys. {\bf 33}(2003)1033.

\item{[22]} Xulu, S.S.: Mod. Phys. Lett. {\bf A15}(2000)1511;
Astrophys. and Space Science {\bf 283}(2003)23.

\item{[23]} Bringley, T.: Mod. Phys. Lett. {\bf A17}(2002)157.

\item{[24]} Som, M.M. and Raychaudhuri, A.K.: Proc. R. Soc. {\bf
A304}(1968)81;

Banerjee A. and Banerji, S.: J. Phys. A: Gen. Phys. {\bf
1}(1968)188.

\item{[25]} Raychaudhuri, A.K. and Thakurta, S.N.: Phys. Rev. {\bf
D22}(1980)802.

\item{[26]} Reboucas, M.J. and Tiomno, J.: Phys. Rev. {\bf
D28}(1985)1251.

\item{[27]} Reboucas, M.J. and Aman, J.E.: J. Math. Phys. {\bf
28}(1987)888.

\item{[28]} Reboucas, M.J. and Teixeira, A.: J. Math. Phys. {\bf
33}(1992)2885.

\item{[29]} Reboucas, M.J. and Aman, J.E.: J. Math. Phys. {\bf
40}(1999)4011.

\item{[30]} Misner, C.W., Thorne, K.S. and Wheeler, J.A.: {\it
Gravitation} (W.H. Freeman, New York, 1973)603.

\item{[31]} Cooperstock, F.I. and Sarracino, R.S.: J. Phys. A.:
Math. Gen. {\bf 11}(1978)877.

\item{[32]} Bondi, H.: {\it Proceedings of the Royal Society of
London} {\bf A427}(1990)249.

\item{[33]} Yang, I.C. and Radinschi, I.: gr-qc/0309130.

\end{description}

\end{document}